\documentstyle[multicol,prl,aps,amsfonts,]{revtex}
\def\bm#1{\mbox{\boldmath $#1$}}

\newcommand{\bra}[1]{\mbox{$\langle{#1}|$}}
\newcommand{\ket}[1]{\mbox{$|{#1}\rangle$}}

\begin{document}

\title{
Modification of relative entropy of Entanglement 
\thanks{Supported by the National Natural Science Foundation of China under Grant No. 69773052 and the Fellowship of China Academy of Sciences}}
\author{An Min WANG$^{1,2}$}
\address{
CCAST(World Laboratory) P.O.Box 8730, Beijing 100080$^1$\\
and Laboratory of Quantum Communication and Quantum Computing$^2$, 
Department of Modern Physics\\
University of Science and Technology of China, 
P.O. Box 4, Hefei 230027, People's Republic of China$^3$}

\maketitle

\begin{abstract}
{We present the modified relative entropy of entanglement (MRE) in order to both improve the computability for the relative entropy of entanglement and avoid the problem that the entanglement of formation seems to be greater than entanglement of distillation. For two qubit system we derive out an explicit and ``weak" closed expression of MRE that depends on the pure state decompositions in the case of mixed states. For more qubit system, we obtain an algorithm to calculate MRE in principle.  MRE significantly improves the computability of relative entropy of entanglement and decreases the dependence and sensitivity on the pure state decompositions. Moreover it is able to inherit most of the important physical features of the relative entropy of entanglement. In addition, a kind of states, as an extension of Werner's states, is discussed constructively.}

\smallskip
{\noindent}PACS: 03.65.Ud  03.67.-a  \vfill
\end{abstract}

\begin{multicols}{2}

The entanglement is a vital feature of quantum information. It has important applications for quantum communication and quantum computation, for example, quantum teleportation \cite{Bennett1}, massive parallelism of quantum computation \cite{Ekert}, decoherence in quantum computer \cite{QC} and quantum cryptographic schemes \cite{QCY}. In
the existing measures of entanglement, the entanglement of formation (EF) $E_{EF}$ \cite{Bennett} and the relative entropy of entanglement (RE) $E_{RE}$ \cite{Vedral1} are defined by
\vskip -0.1in
\begin{eqnarray}
E_{EF}(\rho_{AB})&=&\min_{\{p_i,\rho^i\}\in{\cal{D}}}\sum_{i} p_iS(\rho_B^i),\label{EF}\\
E_{RE}(\rho_{AB})&=&\min_{\rho^R_{AB}\in {\cal{R}}}S(\rho_{AB}||\rho^{\rm R}_{AB}),\label{RE}
\end{eqnarray}
where ${\cal{D}}$ in Eq.(\ref{EF}) is a set that includes all the possible decompositions of pure states $\rho=\sum_i p_i\rho^i$, and ${\cal{R}}$ in Eq.(\ref{RE}) is a set that includes all the disentangled states. 
Note that $\rho_B^i={{\rm Tr}_A\rho^i}$ is the reduced density matrix of $\rho^i$, $S(\rho)$ is von Neumann entropy of $\rho$, $S(\rho||\rho^{\rm R})={\rm Tr}(\rho\log\rho-\rho\log\rho^{\rm R})$ is the quantum relative entropy and $\rho^{\rm R}$ can be called the relative density matrix. 

For a pure state in a bi-party system EF is an actually standard measure of entanglement. For an arbitrary state of two qubits, EF is also widely accepted \cite{Wootters}. However, there is a surprised result recently, \cite{Vedral1} that is, it seems that EF is greater than the entanglement of distillation (DE) in the case of mixed state. If it is so, how to calculate RE will be important because RE is thought of an upper bound of DE in the case of mixed states \cite{Vedral1}. 
RE also appears promising by a series of the interesting results \cite{Vedral2}. However, its advantages suffers from the difficulty in computation. The set ${\cal{R}}$ is so large that one can not sure when the minimumizing procedure is finished. In this letter, in order to avoid the problem that EF seems to be greater than DE and improve the computability of RE, we present a modified relative entropy of entanglement (MRE) as a possible upper bound of DE, through redefining a suitable relative density matrix first for the given pure states, and then for mixed states in terms of the physical idea to choose a minimum pure state decomposition such as EF. The physical intuition to propose MRE is original from organically combining the advantages of EF for the pure states and RE for the mixed states and avoiding their shortcomings as possibly. 

How to calculate RE is also interesting even if in the simple enough case of two qubits except for one can prove that for an arbitrary state of two qubits EF calculated by Wootters' method is not greater than DE. If this exception happens, RE is only important for a bigger system more than two qubits and then we have to study the relevant problem and further development based this letter. That is, we will extend MRE to multi-party systems naturally and overcome the difficulty that one only can discuss RE qualitatively for multi-party systems \cite{My1}. 
  
For our purpose, let's first give out three lemmas.

{\it Lemma one}.\ For two qubits, the polarization vectors $\bm{\xi}_{A}$ and $\bm{\xi}_B$ corresponding to the reduced matrices $\rho_{\{A,B\}}=\frac{1}{2}(\bm{1}+\bm{\xi}_{\{A,B\}}\cdot\bm{\sigma})$ read:
\begin{equation}
\bm{\xi}_A={\rm Tr}(\rho \bm{\sigma}\otimes I),\quad\bm{\xi}_B={\rm Tr}(\rho I\otimes\bm{\sigma}),
\end{equation}
where $\bm{\sigma}$ is the Pauli spin matrix.

{\it Lemma Two}.\ For the pure state of two qubits, there are the relations between  $\bm{\xi}_{A}$ and $\bm{\xi}_B$
\begin{equation}
\xi^i_A=4\sum_{j=1}^3a_{ij}\xi^j_B,\quad 4\sum_{i=1}^3\xi^i_Aa_{ij}=\xi^j_B,
\end{equation}
where $a_{ij}$ is the expanding coefficients
\begin{equation}
\rho=\sum_{\mu,\nu=0}^{3} a_{\mu\nu}\sigma_\mu\otimes\sigma_\nu,
\end{equation}
in which $\sigma_0$ is the identity matrix. In the case of a pure state $\ket{\psi}=a\ket{00}+b\ket{01}+c\ket{10}+d\ket{11}$, it follows that
\begin{equation}
\bm{\xi}^2=\bm{\xi}^2_A=\bm{\xi}^2_B=1-4|ad-bc|^2.
\end{equation}

{\it Lemma three}.\ If the relative density matrix in its eigenvector decomposition is:
\begin{equation}
\rho^{\rm R}=\sum_{\alpha}\lambda_\alpha\rho_\alpha^{\rm R}=\sum_{\alpha}\lambda_\alpha\ket{v_\alpha^{\rm R}}\bra{v_\alpha^{\rm R}},\label{ED}
\end{equation}
where $\lambda_\alpha$ is taken over all the eigenvalues and the eigen density matrices are assumed to be orthogonalized and idempotent without loss of generality,  
thus the relative entropy can be written as
\begin{eqnarray}
S(\rho||\rho^{\rm R})&=&-S(\rho)-\sum_{\alpha}\log \lambda_\alpha {\rm Tr}(\rho\rho_\alpha^{\rm R})\\ \nonumber
&=&-S(\rho)-\sum_{\alpha}\log \lambda_\alpha \bra{v_\alpha^{\rm R}}\rho\ket{v_\alpha^{\rm R}}.\label{CRE}
\end{eqnarray}

From the definition of the polarization vectors, Lemma one is easy to get. To prove Lemma two, it is used the fact that $\rho^2=\rho$ for the pure state and Lemma one. It is easy to prove Lemma three by the simple computation in quantum mechanics. Lemma three implies that the key to calculate RE is to seek an appropriate relative density matrix $\rho^{\rm R}$ and to find out its all the eigenvalues. 

Now we can formulate the basic theorems of this letter.

{\it Theorem one}. In the case of the pure state $\rho^{\rm P}$ of two qubits, the relative density matrix of RE can be taken as
\begin{eqnarray}
R(\rho^P)&=& \frac{1+|\bm{\xi}|\;}{2}\frac{1}{2}\left(I+\frac{\bm{\xi}_A}{|\bm{\xi}|} \cdot \bm{\sigma}\right)\otimes \frac{1}{2}\left(I+\frac{\bm{\xi}_B}{|\bm{\xi}|} \cdot \bm{\sigma}\right)\\ \nonumber
& &+\frac{1-|\bm{\xi}|\;}{2}\frac{1}{2}\left(I-\frac{\bm{\xi}_A}{|\bm{\xi}|} \cdot \bm{\sigma}\right)\otimes \frac{1}{2}\left(I-\frac{\bm{\xi}_B}{|\bm{\xi}|} \cdot \bm{\sigma}\right).
\end{eqnarray}
For the maximum entanglement state 
\begin{equation}
\ket{\Phi^\pm}=\frac{1}{\sqrt{2}}(\ket{00}\pm\ket{11},\quad \ket{\Psi^\pm}=\frac{1}{\sqrt{2}}(\ket{01}\pm\ket{10},
\end{equation}
because that $|\bm{\xi}|=0$, it should be
\begin{eqnarray}
R(\Phi^\pm)&=&\frac{1}{2}\left( \frac{1}{2}\left(I+\sigma_3\right)\otimes \frac{1}{2}\left(I+\sigma_3\right)\right)\\ \nonumber
& &+\frac{1}{2} \left(\frac{1}{2}\left(I-\sigma_3\right)\otimes \frac{1}{2}\left(I-\sigma_3\right)\right),\\
R(\Psi^\pm)&=&\frac{1}{2}\left( \frac{1}{2}\left(I+\sigma_3\right)\otimes \frac{1}{2}\left(I-\sigma_3\right)\right)\\ \nonumber
 & &+\frac{1}{2} \left(\frac{1}{2}\left(I-\sigma_3\right)\otimes \frac{1}{2}\left(I+\sigma_3\right)\right).
\end{eqnarray}
That RE are calculated in terms of $R(\rho^{\rm P})$ for a pure state is equal to EF.

In order to prove this theorem, we, in Eq.(\ref{RE}), choose such a subset of ${\cal{R}}$ that $\rho^\alpha$ is purely separable as 
$\rho^\alpha=\rho_A^\alpha\otimes\rho_B^\alpha$. 
For simplicity, only consider the case with two qubits. Because that the eigen density matrix is pure, $\rho_A^\alpha$ and $\rho_B^\alpha$ have to be pure. While the $2\times 2$ density matrix of the pure state can be written as $
\rho_{\{A,B\}}^\alpha=\frac{1}{2}(1+\bm{\eta}_{\{A,B\}}^\alpha\cdot \bm{\sigma})$, where $|\bm{\eta}_{\{A,B\}}^\alpha|=1$. Thus, 
from Lemma three it follows that
\begin{eqnarray}
S(\rho||\rho^{\rm R})&=&-S(\rho)-\sum_{\alpha}\log \lambda_\alpha \sum_{\mu,\nu=0}^3 \eta^\alpha_A{}_\mu a_{\mu\nu}\eta^\alpha_B{}_\nu\\ \nonumber
&=&-S(\rho)-\sum_{\alpha}\omega^\alpha \log\lambda_\alpha, 
\end{eqnarray}
where $\omega^\alpha=\displaystyle \sum_{\mu,\nu=0}^3 \eta^\alpha_A{}_\mu a_{mu\nu}\eta^\alpha_A{}_\nu$ and $\eta_{\{A,B\}}=(1,\bm{\eta}_{\{A,B\}})$. 
Because of the orthogonal property among the different $\rho^\alpha$, we can choose
$\bm{\eta}_A^1=-\bm{\eta}_A^3=\bm{k}$,  $\bm{\eta}_A^2=-\bm{\eta}_A^4=\bm{m}$ and $\bm{\eta}_B^1=-\bm{\eta}_B^2=\bm{\eta}_B^3=-\bm{\eta}_B^4=\bm{n}$. In order to  calculate the minimum value of the relative entropy, one has to find the partial derivatives of all the variables, set them to zero to form a equation system, and then solve this equation system. However, it doesn't exist. So we only find the extreme surface fixing all the eigenvalues of $\rho^{\rm R}$. 
It is easy to verify that 
$\lambda_\alpha=\omega_\alpha$ gives out the minimum surface. Thus,  
the minimum relative entropy is
\begin{equation}
S(\rho||\rho^{\rm R})=-S(\rho)-\sum_{\alpha}\omega^\alpha \log\omega_\alpha, 
\end{equation}
Again substituting into the chosen $R(\rho^{\rm P})$ in Theorem one, in terms of all of lemmas, it is obtained immediately
\begin{equation}
E_{RE}(\rho^{\rm P})=S(\rho^{\rm P}||R(\rho^{\rm P}))=S(\rho_{\{A,B\}}^{\rm P})=E_{EF}(\rho^{\rm P}).
\end{equation}

For the mixed state, if theorem one is extended directly, we will find the result is not satisfied. So, in terms of the physical idea of EF to deal with the case of the mixed state, we obtain 

{\it Definition}.\ For a pure state $\rho^{\rm P}$ and a mixed state $\rho^{\rm M}$, MRE is defined respectively as
\begin{eqnarray}
E_{MRE}(\rho^{\rm P})&=&S(\rho^{\rm P}||R(\rho^{\rm P}))=E_{EF}(\rho^{\rm P})\label{MREforP}, \\
E_{MRE}(\rho^{\rm M})&=&\min_{\{p_i,\rho^i\}\in{\cal{D}}} S\left(\rho^{\rm M}||\sum_{i}p_iR(\rho^i)\right)\label{MREforM}\\ \nonumber
&=&\min_{\{p_i,\rho^i\}\in{\cal{D}}} S\left(\rho^{\rm M}||R^{\rm M}\right),
\end{eqnarray}
where $R(\rho^{\rm P})$ is such a relative density matrix corresponding to the pure state $\rho^{\rm P}$ that Eq.(\ref{MREforP}) is satisfied and $R(\rho^{\rm P})$ is an disentangled density matrix. In Eq.(\ref{MREforM}), the minimum is taken over the set ${\cal{D}}$ that includes all the possible decompositions of pure states  $\rho=\sum_i p_i\rho^i$, and $R^{\rm M}=\sum_{i}p_iR(\rho^i)$ is the total relative density matrix for a mixed state, while $R(\rho^i)$ is found out so that Eq.(\ref{MREforP}) is valid for the pure state $\rho^i$. In particular, for two qubits, $R(\rho^i)$ is chosen by Theorem one. 

{\it Theorem two}\ \ Modified relative entropy of entanglement (MRE) always satisfies:
\begin{equation}
E_{MRE}(\rho)\leq E_{EF}(\rho).
\end{equation}
When $\rho$ is a pure state, the equality is valid. 

It is easy to prove it in terms of the joint convexity of the relative entropy
\begin{equation}
S(\sum_i p_i\rho^i||\sum_i p_i R(\rho^i))\leq \sum_i p_iS(\rho^i||R(\rho^i))
\end{equation}
and the definition of $E_{EF}$ in Eq.(\ref{EF}). Obviously for a pure state, MRE is equal to RE and EF. The relative density matrix $R$ for MRE in a given pure state can be defined by this relation, that is solving $S(\rho^{\rm P}||R)=S(\rho_B^{\rm P})=E_{EF}(\rho^{\rm P})$ to find $R$. For the case of mixed state, one first finds the relative density matrix $R(\rho_i)$, in which $\rho_i$ belong to a pure state decomposition, by solving $S(\rho^i||R(\rho^i))=S(\rho_{B}^{i})=E_{EF}(\rho^i)$. Then, one can write the total relative density matrix for a mixed state as $R^{\rm M}=\sum_{i}p_iR(\rho^i)$ and calculate the corresponding relative entropy. The last, MRE is obtained by taking the minimum one among these relative entropies. Therefore we obtain a definite and constructive algorithm to calculate MRE as we have had in the calculation of EF. For two qubit systems, we have successfully obtained the explicit and general expression of the relative density matrix in a pure state or a mixed state with any given decomposition. MRE for two qubit systems can be easier calculated because the first step in our algorithm is largely simplified. 
For more than two qubits, we do not give clearly an explicit expression of the relative density matrix for a pure state in this paper, the first step needs more computations, but our algorithm still works in principle. This is because that from $S(\rho_i^{\rm P}||R)=S(\rho_{B}^{i})=E_{EF}(\rho^i)$ to find $R(\rho^i)$ can be done within finite steps for a given pure state in general except for the solution $R(\rho^i)$ does not exist. The exception is impossible because this implies that RE for the pure state $\rho_i$ has no a relative density matrix ($S(\rho_{B}^{i})=\min S(\rho_i||\rho^R)=E_{RE}(\rho^i)$). It must be emphasized that our method is to calculate MRE but not EF, and the comparison between MRE and EF is given in their algorithm but not in their physical significance and requirement. In other words, EF and MRE can not be replaced each other, our method and the computational method of EF, for example Wootter's method, can not be replaced each other. In above sense, MRE avoids the difficulty of RE to find the relative density matrix from an infinite large set of disentangled states and so improve the computability of RE. Moreover, in our recent paper \cite{My1}, we have given an explicit expression of the relative density matrix for $n$-party systems (restricted to qubits). 

In the case of mixed states, obviously RE$\leq$MRE$\leq$EF. Noting the fact that both RE and MRE are defined by the relative entropy and satisfy Theorem two, we think that MRE is able to inherit most of important physical features of RE if these features of RE are given and proved in terms of the fact stated above as well as some mathematical skills \cite{Vedral1,Vedral2}. For example, the properties under local general measurement (LGM) and classical communication (CC) can be proved by using of the similar methods in refs.\cite{Vedral1,Vedral2,My2} at least for two qubit systems. 
We have seen that MRE is the function of the polarization vectors of the reduced density matrices of the decomposition density matrices for two qubits. Thus, EF and RE as well as MRE all belong to a kind of the generalized measures of entanglement proposed by \cite{My2}, and the generalized measures of entanglement with all the known properties as a good measure are proved there. Two known main measures of entanglement are related together by the polarization vectors of the reduced density matrix. As to the properties in two qubit systems, such as its range is $[0,1]$, its maximum value $1$ corresponds to the maximally entangled states and its minimum value $0$ corresponds to the mixture of the disentangled states, can be directly and easily obtained from the definition of MRE. 
For two qubits, the relative density matrix is a function of the polarized vectors $\bm{\xi}_A^i,\bm{\xi}_B^i$, and $\bm{\xi}_A^i,\bm{\xi}_B^i$ are functions of the decomposition density matrices $\rho_i$. Thus, MRE is just a compound function of the decomposition density matrices $\rho_i$. Of course, it is not a good property that a measure of entanglement depends on the possible decompositions because it is not very easy to find all the elements of ${\cal{D}}$. But this exists in all the known measures of entanglement either. MRE has significantly improvement in this aspect for some kinds of states. For example, the state $D$ has two pure state decompositions
\vskip -0.1in
\begin{eqnarray}
D&=&\frac{1}{2}\left(\ket{00}\bra{00}+\ket{11}\bra{11}\right)\\
&=&\frac{1}{2}\left(\ket{\Phi^+}\bra{\Phi^+}+\ket{\Phi^-}\bra{\Phi^-}\right)
\end{eqnarray}
which respectively correspond to the minimum and maximum decompositions in the calculation of EF. But two decompositions have the same relative density matrices in the calculation of MRE. That is, both of them are the minimum for MRE and can be used to calculate MRE. This means that the minimum decomposition(s) to calculate MRE is (are) not the same as the minimum decomposition(s) to calculate EF in general. We can verify that this advantage is kept for Werner's state \cite{Werner} 
\vskip -0.1in
\begin{eqnarray}
W&=&F\ket{\Psi^-}\bra{\Psi^-}+\frac{1-F}{3}(\ket{\Psi^+}\bra{\Psi^+}\\ \nonumber
& &+\ket{\Phi^+}\bra{\Phi^+}+\ket{\Phi^-}\bra{\Phi^-})
\end{eqnarray}
Its relative density matrix reads $R(W)_{ij}(i\neq j)=0, R(W)_{ii}=\{(1-F)/3,(1+2F)/6,(1+2F)/6,(1-F)/3\}$. Thus, it is easy to get
\vskip -0.1in
\begin{eqnarray}
S(W||R(W))
&=&F\log F+\frac{1-F}{3}\log\left(\frac{1-F}{3}\right)\nonumber\\
& &-\frac{1+2F}{3}\log\left(\frac{1+2F}{6}\right),\label{MREforWS}
\end{eqnarray}
It correctly gives out MRE of Werner state. In fact, from Eq.(\ref{MREforWS}) it follows that when $F=1/4$ Werner's state is disentangled, and when $F=1$ Werner's state has the maximum entanglement. If we take 
\begin{eqnarray}
W&=&\frac{4F-1}{3}\ket{\Psi^-}\bra{\Psi^-}+\frac{1-F}{3} I\quad \left(F\geq\frac{1}{4}\right)\\
W&=&\frac{1-4F}{3}\ket{\Psi^+}\bra{\Psi^+}+\frac{1-F}{3}(\ket{00}\bra{00}+\ket{11}\bra{11})\\ \nonumber
& &+F(\ket{01}\bra{01}+\ket{10}\bra{10})\quad \left(F\leq\frac{1}{4}\right),
\end{eqnarray}
the relative density matrix does not change and so the result is the same. This implies that MRE can decrease the dependence and sensitivity on the pure state decompositions at least for some interesting states. We can more easily find an adequate pure state decomposition in the calculation of MRE than do this in the calculation of EF. 

If we extend Werner's state to a new kind of states
\begin{equation}
W_{\rm E}=\sum_{i=1}^4 b_i \ket{{\rm{B}_i}}\bra{{\rm B}_i}+\sum_{i=1}^4 c_i\ket{i}\bra{i},\label{EWS}
\end{equation}
where $\ket{\rm{B}_i}$ are four Bell states $\ket{\Phi^\pm},\;\ket{\Psi^\pm}$ and $\ket{i}$ are $\ket{00},\;\ket{01},\;\ket{10},\;\ket{11}$ respectively. Note that $\sum_{i=1}^4 (b_i+c_i)=1$ 
and all of them are positive. We can find that MRE also depends on the pure state decompositions. For the simplicity, consider the state
\begin{equation}
\rho=\lambda\ket{\Phi^+}\bra{\Phi^+}+(1-\lambda)\ket{00}\bra{00}.
\end{equation}
Its eigen decomposition is
\begin{eqnarray}
\rho&=&v_-\ket{V^-}\bra{V^-}+v_+\ket{V^+}\bra{V^+},\\  \nonumber
v_\pm&=&\frac{1}{2}\left(1\pm\sqrt{1-2\lambda(1-\lambda)}\right).
\end{eqnarray}
Two decompositions lead to the different the relative density matrices and relative entropies. Therefore, this implies that the minimum procedure is not unnecessary for the modified relative entropy of entanglement in general. It seems to us, it is interesting to give a good algorithm that can find all the elements of the set of the pure state decompositions ${\cal{D}}$. This is still an open question. But for a kind of mixed states with the form of the extension of Werner's state (\ref{EWS}), we can find that its definition is an adequate minimum decomposition (its reliability has not been strictly proved in mathematics, but we have not found a counterexample). This conclusion can be obtained perhaps because MRE of this kind of states is not very sensitive to the pure state decompositions. Our method is first to choose a minimum decomposition among all of decomposition that we can find, then check the result by the particular points (disentangled and maximum entangled) and repeat this process up to the case we can not continue it. Furthermore, we carry out some numerical checking. Thus, MRE of the extended Werner state can be written as
\begin{eqnarray}
E_{MRE}(W_{\rm E})&=&\sum_{\alpha} v_\alpha\log v_\alpha \\ \nonumber & &-\frac{1}{2}(b_1+b_2 +2c_1)\log\frac{1}{2}(b_1+b_2 +2c_1)\\ \nonumber
& &-\frac{1}{2}(b_3+b_4 +2c_2)\log\frac{1}{2}(b_3+b_4 +2c_2)\\ \nonumber & &-\frac{1}{2}(b_3+b_3 +2c_4)\log\frac{1}{2}(b_3+b_4 +2c_3)\\ \nonumber
& & -\frac{1}{2}(b_1+b_2 +2c_4)\log \frac{1}{2}(b_1+b_2 +2c_4),
\end{eqnarray}
where the eigenvalues of $W_{\rm E}$ are
\begin{eqnarray}
v_1&=&\frac{1}{2}(b_1+b_2+c_1+c_4-\sqrt{(b_1-b_2)^2+(c_1-c_4)^2}),\\ \nonumber
v_2&=&\frac{1}{2}(b_1+b_2+c_1+c_4+\sqrt{(b_1-b_2)^2+(c_1-c_4)^2}),\\ \nonumber
v_3&=&\frac{1}{2}(b_3+b_4+c_2+c_3-\sqrt{(b_3-b_4)^2+(c_2-c_3)^2}),\\ \nonumber
v_4&=&\frac{1}{2}(b_3+b_4+c_2+c_3+\sqrt{(b_3-b_4)^2+(c_2-c_3)^2}).
\end{eqnarray}
Based on Peres's condition, \cite{Peres} we can calculate the eigenvalues of partial transpose of the extended Werner state \cite{My4} and obtain the condition that $W_E$ is separable 
\begin{eqnarray}
(b_1 +b_2)^2&\geq& (b_3-b_4)^2-4c_1c_4,\\ \nonumber
(b_3+b_4)^2&\geq& (b_1-b_2)^2-4c_2c_3.
\end{eqnarray}

In conclusion, MRE can be useful based on four evidences. One is that MRE is a possible upper bound of entanglement of distillation  such as RE, the second is MRE improves the compatibility of RE, the third is that MRE significantly decrease the dependence and sensitivity on the pure state decompositions at least for some interesting states, and the last is that MRE can be extended to multi-party systems naturally\cite{My1}. This research is on progressing.

\end{multicols}
\end{document}